\documentclass{elsart5p}
\usepackage{graphicx,natbib,amssymb,lineno}

\begin{document}

\begin{frontmatter}
\title{\textbf{Emission Heights of Coronal Bright Points on Fe XII Radiance Map}}

\author[China]{\textbf{H. Tian}},
\author[China]{\textbf{C.-Y. Tu}\corauthref{cor}},
\corauth[cor]{Corresponding author.} \ead{chuanyitu@pku.edu.cn}
\author[China]{\textbf{J.-S. He}},
\author[Germany]{\textbf{E. Marsch}}

\address[China]{Department of Geophysics, Peking University, Beijing 100871, China}
\address[Germany]{Max-Planck-Institut f$\ddot{u}$r Sonnensystemforschung, 37191 Katlenburg-Lindau, Germany}

\begin{abstract}
The study of coronal bright points (BPs) is important for
understanding coronal heating and the origin of the solar wind.
Previous studies indicated that coronal BPs have a highly
significant tendency to coincide with magnetic neutral lines in the
photosphere. Here we further studied the emission heights of the BPs
above the photosphere in the bipolar magnetic loops that are
apparently associated with them. As BPs are seen in projection
against the disk their true emission heights are unknown. The
correlation of the BP locations on the Fe XII radiance map from EIT
with the magnetic field features (in particular neutral lines) was
investigated in detail. The coronal magnetic field was determined by
an extrapolation of the photospheric field (derived from 2-D
magnetograms obtained from the Kitt Peak observatory) to different
altitudes above the disk. It was found that most BPs sit on or near
a photospheric neutral line, but that the emission occurs at a
height of about 5 Mm. Some BPs, while being seen in projection,
still seem to coincide with neutral lines, although their emission
takes place at heights of more than 10 Mm. Such coincidences almost
disappear for emissions above 20 Mm. We also projected the upper
segments of the 3-D magnetic field lines above different heights,
respectively, on to the tangent x-y plane, where x is in the
east-west and y in the south-north direction. The shape of each BP
was compared with the respective field-line segment nearby. This
comparison suggests that most coronal BPs are actually located on
the top of their associated magnetic loops. Finally, we calculated
for each selected BP region the correlation coefficient between the
Fe XII intensity enhancement and the horizontal component of the
extrapolated magnetic field vector at the same x-y position in
planes of different heights, respectively. We found that for almost
all the BP regions we studied the correlation coefficient, with
increasing height, increases to a maximal value and then decreases
again. The height corresponding to this maximum was defined as the
correlation height, which for most bright points was found to range
below 20 Mm.
\end{abstract}

\begin{keyword}
Coronal bright points; Magnetic loops; Correlation height
\end{keyword}
\end{frontmatter}

\section{\large Introduction}
\label{intro}

Coronal bright points (BPs) were first observed in soft X-ray images
\citep{Vaiana70}, and later found to be also visible with enhanced
intensity at EUV and radio wavelengths. This enhanced emission
feature is 30$^{\prime\prime}$-40$^{\prime\prime}$ in size, often
with a bright core of 5$^{\prime\prime}$-10$^{\prime\prime}$
\citep{Madjarska03}.The lifetime of EUV bright point ranges from 5
hours to 40 hours, with an average of 20 hours \citep{Zhang01},
while the average lifetime of an X-ray bright point is 8 hours as
determined by Skylab X-ray observations \citep{Golub74}.
\citet{Braj04} used coronal BPs as tracers to analyze solar
differential rotation, and concluded that the average height of BPs
in different 10-degree-latitude bins differs a lot (from about 5000
km to 22000 km), but on average is 8000 km to 12000 km above the
photosphere.

Coronal bright points are associated with regions of mixed-polarity
magnetic flux of the magnetic network \citep{Webb93, Falconer98,
Wilhelm00, Xia03}. And most BPs are likely to be associated more
with the cancellation of magnetic features than their emergence
\citep{Webb93}. Recent observations based on EIT/SOHO and TRACE
confirmed that BP is strongly correlated with the evolution of the
underlying bipolar magnetic region \citep{Brown01, Madjarska03,
Ugarte04}. Some BPs are observed to lie at the base of polar plumes
in coronal holes and may eject mass into the solar wind
\citep{Ahmad78}. However, SUMER/SOHO observations indicated that BP
regions correspond to no or small outflow, and it was concluded that
the fraction of the total mass flux contributed by BPs to the solar
wind is negligible \citep{Wilhelm00, Madjarska03, Xia03}.
Spectroscopic analyses based on HRTS, SUMER, and CDS revealed the
presence of small-scale transient brightenings within the bright
point, but no relation was found between the bright point and the
explosive events \citep{Moses94, Madjarska03, Ugarte04}.

It was suggested that BPs may result from the interaction between
two magnetic fragments with opposite polarities, which can lead to
magnetic reconnection and local heating of the plasma
\citep{Priest94, Parnell94, Neukirch97, Von06}. The model proposed
by \citet{Parnell94} also suggested that BPs lie near the location
where magnetic features cancel but not necessarily directly above
it. Recently, \citet{Buchner04a} and \citet{Buchner04b} simulated
the consequent formation of non-force-free current sheets in
chromosphere and corona. Their model results suggest that the
current sheets induced by photospheric motion supply the energy for
the energization of BP.

The magnetic field used in the above cited observational analyses
was the photospheric field. However, a BP is a EUV radiation
phenomenon that occurs in the corona. So, a comparison should be
made with the coronal magnetic field. Since the magnetic field above
the photosphere can still not be obtained through direct
observation, an extrapolation from a photospheric magnetogram is
required, which will usually provide a good approximation to the
real coronal field. For a review of this subject, see
\citet{Wiegelmann02}.

In our present study we used the force-free-field extrapolation
method proposed by \citet{Seehafer78} to extrapolate the field based
on a Kitt Peak photospheric magnetogram to the corona. Then we
combined this extrapolated magnetic field with the Fe XII radiance
map from EIT, in order to study the relation of BPs with magnetic
neutral lines and projections of magnetic loops. From these
comparisons the height range of the BP locations could be roughly
estimated.

\section{\large Data analysis}

\begin{figure}
\centering
\includegraphics[width=7.6cm]{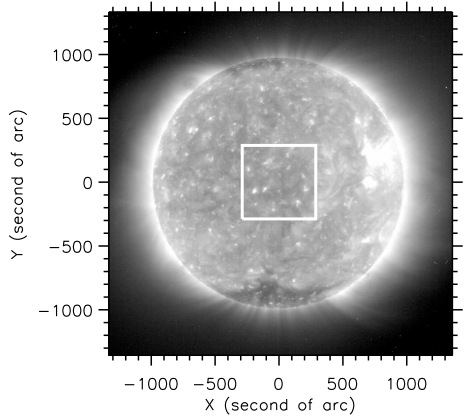}
\caption{Fe XII (19.5 nm) intensity image from an observation made
by SOHO/EIT at 19:13 UT on 16 March 1997. The white square with a
width of 0.6 solar radii on the disk center is the area which we
chose for our study.} \label{fig1}
\end{figure}

We used a Fe XII (19.5 nm) filtergram from the observation that
SOHO/EIT made at 19:13 UT on 16 March 1997. The pixel size of this
coronal image is 2.6 seconds of arc, which is high enough to study
coronal BPs. Our selected area of the image is a square centered on
the solar disk, has a width of 0.6 solar radii and obviously does
not include an active region. This area is shown in Figure
\ref{fig1} as a white rectangle.

The spatial filtering method described by \citet{Falconer03} was
used to produce the image of BPs. For the Fe XII intensity image, we
chose two squares centered on each pixel, an outer square with a
size of about two thirds of a network cell (about 21000 km, or 11
pixels), and an inner square that was about one third of a network
cell wide (about 9000 km, or 5 pixels). For each pixel, the
background intensity was defined as follows: First the difference
between the integrated intensities over the outer and inner square
was calculated. Then this was divided by the difference between the
areas of the two squares. The resulting ratio was defined as the
background intensity in any given pixel, and hence we got a uniform
background image. By subtracting this background image from the
original one, we obtained an image of the intensity enhancement.
Only contiguous sets of pixels in which the emission was enhanced by
30\% or more above the local background were selected and defined as
BPs, which are shown as dark patches in Figure \ref{fig2}. We also
applied this method to the Fe XII filtergram that was observed six
hours before, and thus found many BPs already existed at that time.
In accordance to \citet{Zhang01} who found lifetime of BPs ranges
from 5 to 40 hours, this result revealed that coronal BPs are not
simply transient events.

We used a Kitt Peak photospheric magnetogram that was obtained on
the same day but about one and a half hours before the time of the
EIT observation. This magnetogram has a pixel size of 1.15 seconds
of arc and was summed over 2$\times$2 pixels to reduce the noise
level, as was discussed by \citet{Falconer98}. Both the magnetic
field and EIT radiation were assumed to remain unchanged during this
period. First the magnetogram was shifted to compensate for the
solar rotation during the two observations.Then we did cross
correlation between the EIT image and the magnetogram and obtained
the optimal match of them. After that we chose the data of the
magnetic field from the same area as the EIT image.

On the basis of the force-free-field assumption, \citet{Seehafer78}
introduced a coronal magnetic field extrapolation technique which
was applied to a finite rectangular segment of the solar atmosphere.
The hypothesis of a force-free field generally can be applied at
heights of about 400 km above the photosphere \citep{Metcalf95}.
This technique has been successfully used to describe the magnetic
field in the transition region and lower corona in the papers by
\citet{Tu05a} and \citet{Tu05b} for a quiet-sun region and coronal
hole and by \citet{Marsch06} for a polar coronal hole. By using this
method, the 3-D magnetic field \textbf{B}(x, y, z) was extrapolated
from the photosphere to the corona. This method is feasible because
a coronal BP is a long-living feature, which lasts for more than
several hours on an EIT image. In our study, the magnetic field was
extrapolated from the photosphere to a height of 80 Mm. Then a
3$\times$3 pixel boxcar smoothing function was applied to the
absolute magnetic flux density at each height for which we made a
data analysis. Those pixels that had a smoothed flux at the
corresponding height below the noise level estimated by the method
described below were masked out. We retained only the polarity data
of the magnetic field in the unmasked pixels, which resulted in a
distribution pattern of magnetic polarities. Figure \ref{fig2} shows
the distribution pattern of magnetic polarities at 5 Mm (left panel)
and 9 Mm (right panel). The regions with positive and negative
magnetic flux density are outlined in blue and red, respectively.
The estimated noise level at a height of \emph{h} Mm, \emph{$n_h$},
was derived from the following formula:
\begin{equation}
\emph{$n_h$}=\frac{\overline{{|\emph{$B_{zh}$}|}}}{\overline{|\emph{$B_{z0}$}|}}\emph{$n_0$}
\end{equation}
where $\overline{{|\emph{$B_{z0}$}|}}$ and
$\overline{{|\emph{$B_{zh}$}|}}$ represent the mean values of the
absolute magnetic flux density at the photosphere and the height of
\emph{h} in Mm, respectively, and \emph{$n_0$} is the noise level of
the Kitt Peak photospheric magnetogram, which is assumed to be 3
Gauss, as discussed in \citet{Falconer98}. At the height of 5 Mm and
9 Mm, this value is 1.256 Gauss and 0.846 Gauss, respectively. In
Figure \ref{fig2}, the two bright points edged by purple curves
(around the coordinates of (70$^{\prime\prime}$,
210$^{\prime\prime}$) in the left panel and (-190$^{\prime\prime}$,
-70$^{\prime\prime}$) in the right panel) will be used as examples
for discussion. We define the direction in which the BP region has
its largest extension as the preferred orientation. We can see that
the preferred orientation of the first BP corresponds to a large
angle (more than 60 degrees) with respect to the polarity-dividing
neutral line. We name this a type I (perpendicular) BP. In contrast,
the preferential orientation of the second variety, which we name
type II (parallel) BP, is almost along the polarity-dividing line.

The real 3-D coronal magnetic field can be reconstructed or
approximated by means of the extrapolated magnetic field. In order
to find the relation between the locations of BPs and their
supporting magnetic loops, we projected the upper segments of the
3-D magnetic field lines above different heights, respectively, onto
the x-y tangent plane, where x is in the east-west direction and y
is in the south-north direction. Figure \ref{fig3} shows these
projections at 5 Mm (left panel) and 9 Mm (right panel). The red
lines represent closed field lines and the blue lines open field
lines. The BP distribution pattern copied from Figure \ref{fig2} was
also re-drawn at 5 Mm (left panel) and 9 Mm (right panel) for
comparison.

\section{\large Results and discussion}

\begin{figure*}
\centering
\includegraphics[width=18cm]{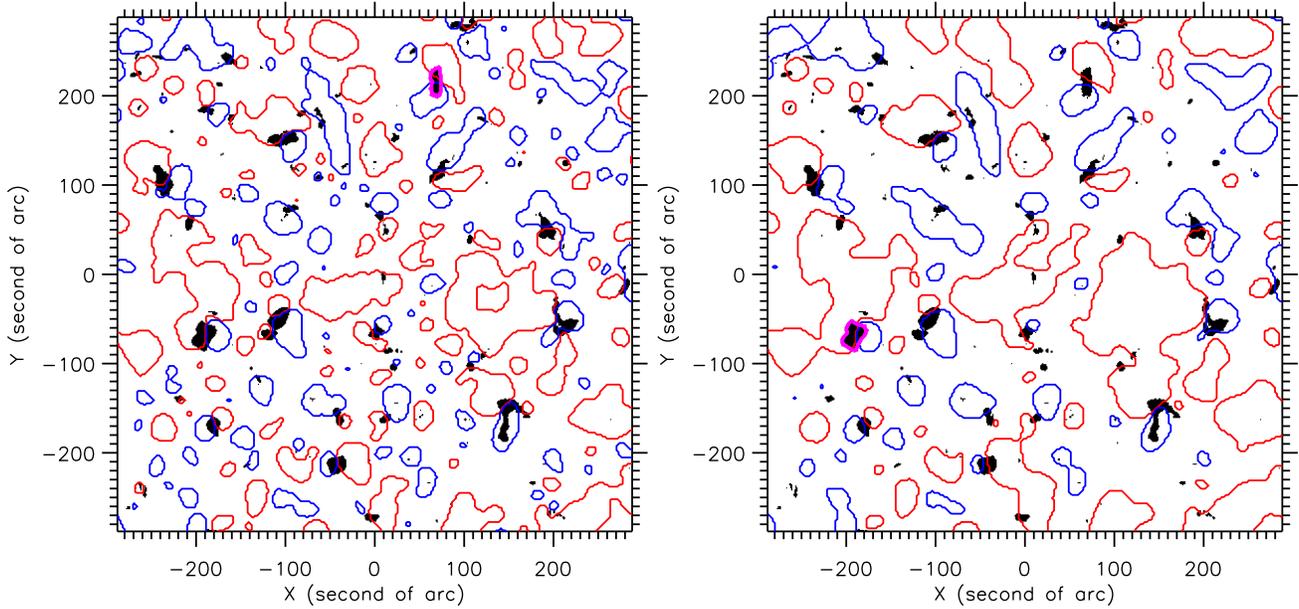}
\caption{The EIT coronal bright points registered are shown on the
map of magnetic polarities at 5 Mm (left panel) and 9 Mm (right
panel). The dark regions represent EIT BPs, the emission of which
was enhanced by more than 30\% above the background. Regions with
positive and negative magnetic flux density are outlined in blue and
red color, respectively. The two BPs edged by purple curves (around
the coordinates of (70$^{\prime\prime}$, 210$^{\prime\prime}$) in
the left panel and (-190$^{\prime\prime}$, -70$^{\prime\prime}$) in
the right panel) are used as examples for the discussion in the
text.} \label{fig2}
\end{figure*}

We found that, based on magnetic field extrapolation to the lower
corona, most BPs are close to the neutral lines determined from
mixed-polarity regions. This finding corroborates the long-known
result that BPs are located close to neutral lines as identified
from magnetic field observations in the photosphere. It is clearly
seen in Figure \ref{fig2} that in considerable number BPs appear
above or close to neutral lines. For example, the BP around the
coordinates (70$^{\prime\prime}$, 210$^{\prime\prime}$) extends from
about 200$^{\prime\prime}$ to 230$^{\prime\prime}$ in y direction.
From the extrapolated magnetic field maps, both at 5 Mm and 9 Mm, we
can see that a bright pattern crosses the nearby neutral line at
about 217$^{\prime\prime}$. For this case we conclude that the BP is
located above a mixed-polarity region, or simply say neutral line.
We examined all 22 BPs seen in Figure \ref{fig2} with sizes lager
than 10$^{\prime\prime}$. We found that more than 14 BPs (64\%)
cover regions between different magnetic polarities based on the
extrapolation to 5 Mm (see Figure \ref{fig2}, left panel). For the
magnetic field extrapolated to 9 Mm (see Figure \ref{fig2}, right
panel) we still found a considerable number of BPs (more than 10,
corresponding to 46\%) that are located above or near neutral lines.
We continued this kind of comparison with the extrapolated field
lines up to a height of 25 Mm. Our results show that some BPs still
coincide with neutral lines based on extrapolation to more than 10
Mm. However, we cannot confirm such coincidence at heights above 20
Mm.

The spatial resolution in this analysis is limited by the low noise
level of the available magnetic field data. The Kitt Peak
magnetogram, which has a spatial resolution of 4.6$^{\prime\prime}$
which is comparable with the data of EIT on SOHO, was the best one
we could get with low noise level. Since the sizes of the BPs we
analyzed in this work are all larger than 10$^{\prime\prime}$, the
major part of the BPs can be determined, and therefore our
conclusion is believable to some extent. The correlation between the
EIT intensity enhancement and the horizontal magnetic field
component, and the non-correlation between the EIT intensity
enhancement and the vertical magnetic field component, both give
further support to the above conclusion (see Paragraph 4-7 in this
section).

\begin{figure*}
\centering
\includegraphics[width=18cm]{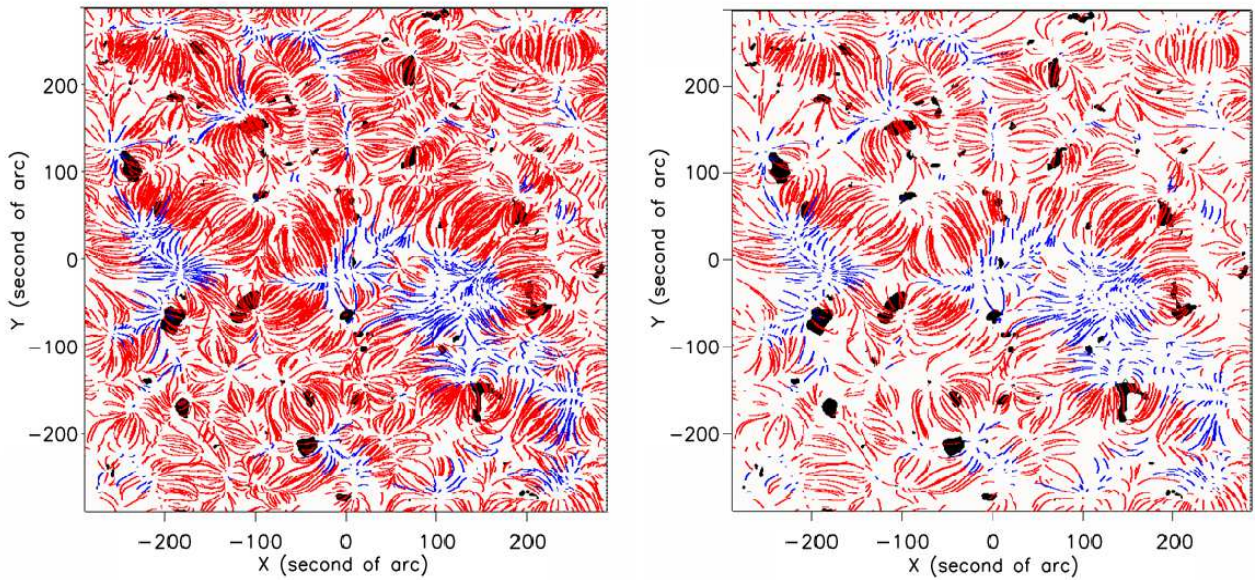}
\caption{Projections of the upper segments of the 3-D magnetic field
lines above 5 Mm (left panel) and 9 Mm (right panel) on the x-y
plane. The red lines represent closed field lines, and the blue
lines correspond to open field lines. Dark regions indicate BPs,
which are the same ones as already shown in Figure \ref{fig2}.}
\label{fig3}
\end{figure*}

Figure \ref{fig3} clearly reveals that nearly all BPs are located in
regions with dense field-line concentrations corresponding to
magnetic loops. Only a few BPs, especially the small ones, are
overlaid on regions with sparse field lines or no loops. One
possible reason may be that the pre-existent loops disappeared
during the time interval between the acquisitions of the EIT coronal
EUV image and the Kitt Peak magnetogram. Another possible reason is
that some loops lower than 5 Mm or 9 Mm just cannot be shown and
discriminated in projection on the images. However, the coincidence
between the locations of magnetic loops and BPs is still very
obvious. This relation suggests that the emission regions giving
rise to most BPs are located on the top of coronal magnetic loops.
From Figure \ref{fig3} we can see the BPs have different shapes,
which may mainly be divided into two types, such as type I
(perpendicular) and type II (parallel) as defined in Section 2. For
the type I BP region, like the one edged by a purple curve in Figure
\ref{fig2} (around the coordinate of (70$^{\prime\prime}$,
210$^{\prime\prime}$) in the left panel), the projections of most
overlying field lines are approximately oriented along the
preferential direction. And for type II BP region, like the one
edged by purple curve in Figure \ref{fig2} (around the coordinate of
(-190$^{\prime\prime}$, -70$^{\prime\prime}$) in the right panel),
most overlying loops are oriented across the preferential direction.

Then we rescaled the EIT image and the extrapolated magnetic field
maps by means of interpolation, thus enforcing them to have the same
pixel size of 1.15$^{\prime\prime}$$\times$1.15$^{\prime\prime}$. In
order to study the emission height of BPs, we calculated the
correlation coefficients between the EIT Fe XII intensity
enhancement and the horizontal component of magnetic field vector at
different heights, respectively, for each BP. If it were located on
the top of a loop, it should be well correlated with the horizontal
component of magnetic field. We found that the value of the
correlation coefficient varied with increasing height. For almost
all the BP regions analyzed, the correlation coefficient increases
to a maximal value before it is decreasing again. The height related
with this maximum was defined as correlation height.

\begin{figure*}
\centering
\includegraphics[width=15cm]{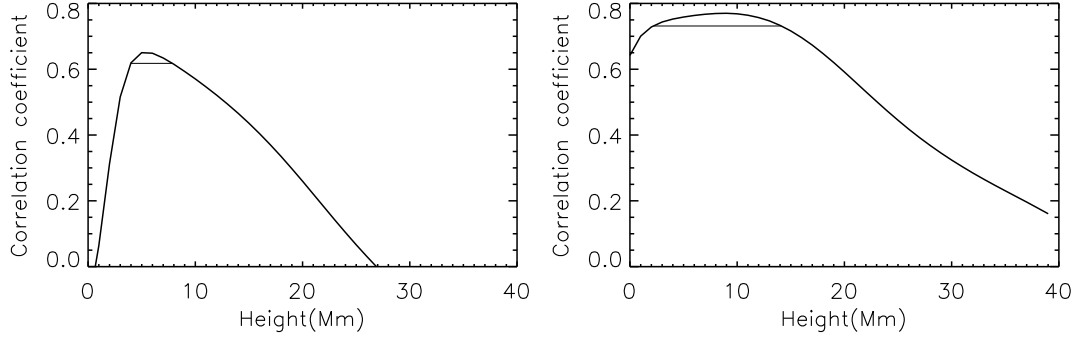}
\caption{Correlation coefficient between the horizontal component of
magnetic field and the square root of the Fe XII intensity
enhancement for the two color-edged BPs shown in Figure \ref{fig2}.
For the first BP (left panel), the maximum correlation coefficient
is 0.65 at 5 Mm. The horizontal bar shows the height range, from
3.99 Mm to 7.85 Mm, in which the coefficient is above 95\% of its
maximum. For the second BP (right panel), the maximum coefficient is
0.77 at 9 Mm. In the height range from 2.20 Mm to 14.11 Mm the
coefficient is above 95\% of its maximum.} \label{fig4}
\end{figure*}

Figure \ref{fig4} shows two examples. For the edged bright point in
Figure \ref{fig2} (left panel), the correlation coefficients at
different heights are shown in Figure \ref{fig2} (left panel). We
found that the correlation coefficient increases to a maximum of
0.65 at 5 Mm before decreasing above this height. The number of data
points in this bright point region is 264, for which the critical
correlation coefficient is 0.12 with 95\% confidence. The horizontal
bar shows a height range from 3.99 Mm to 7.85 Mm, in which the
coefficient is above 95\% of its maximum. Figure \ref{fig4} (right
panel) shows the curve of correlation coefficient for the edged
bright point in Figure \ref{fig2} (right panel). The number of data
points for this calculation is 441, for which a critical value of
0.093 for the linear correlation coefficient is determined with 95\%
confidence. The maximum coefficient is 0.77 at 9 Mm, with a height
range from 2.20 Mm to 14.11 Mm in which the coefficient is above
95\% of its maximum.

For the different BPs, the height ranges obtained by this method are
different. This indicates that BPs do generally not reside at the
same height. Table \ref{table1} lists the correlation heights and
height ranges for 10 large bright points (more than 200 data points,
or with a size of larger than 16$^{\prime\prime}$). We can conclude
that the emission heights of most BPs are below 20 Mm. For all the
selected BPs, the mean value of the correlation height with a
standard error (the standard deviation divided by $\sqrt{10}$) is
10.2$\pm$2.7 Mm. This result is largely consistent with that of
\citet{Braj04}, in which the calculated mean height of the
point-like tracer subtype BP (the subsets of the other two subtypes
of BPs were too small to get reliable results) is 11.6$\pm$2.1 Mm,
when using his interactive method.

We also evaluated the correlation between the Fe XII intensity
enhancement and the vertical magnetic field at different heights,
respectively, for each selected BP. In this case, however, no
obvious correlation was found. This fact proves further that the
emission regions giving rise to most BPs are located on the top of
coronal magnetic loops, where the magnetic field is horizontal. This
picture is consistent with the result of the model suggested by
\citet{Buchner04a} and \citet{Buchner04b}, which implies that the
plasma flow in the photosphere causes the formation of a localized
current sheet in and above the transition region at the position of
the EUV BP, and the dissipation in the current sheet supplies energy
to the BP. Some authors suggested that flares, coronal heating, and
spicules are driven by the same process such as core-field explosion
\citep{Falconer98, Moore99}. If this picture reflects reality, one
would naturally propose that BPs, being indicators of local coronal
heating, can be produced by the same process leading to flares. The
idea that most BP emission is limited to the loop top is consistent
with the observation that hard X-ray emission is often seen at the
top of large flaring loops \citep{Masuda94}. It is possible that
there exists, around the loop top, inverted Y-shaped field lines,
like those proposed in \citet{Masuda94}, and that reconnection
intermittently takes place, thus leading to a bright point. However,
there may be other processes besides reconnection accounting for the
result that the emission in the Fe XII channel of EIT is limited to
the loop apex. If it is not reconnection at the loop top leading to
the BP, then one has to heat up the entire loop and fill it with hot
dense plasma. If then the emission in the Fe XII channel of EIT is
limited to the loop top, this would imply that the top of the loop
supporting a BP has the highest temperature.

It should be pointed out that our results may not be fully
conclusive. Several factors can influence these results. First, the
boundaries of BPs depend on the method we used for identifying the
BPs as described in Section 2. Second, we assumed that the intensity
of BPs and the magnetic field did not change a lot during the time
span between the observations of EIT and Kitt Peak, which is about
one and a half hours. Third, we only apply the correlation method to
BPs with a large size (larger than 16$^{\prime\prime}$). The pixel
size of the final 2-D extrapolated magnetic field map is
2.88$^{\prime\prime}$. To get a sufficiently large number of data
points, we can only apply the correlation method to the relatively
large bright points. This limitation may be overcome only in future
higher-resolution studies.

\section{\large Summary}

We constructed the coronal magnetic field at different heights above
the photosphere by extrapolating photospheric magnetograms obtained
at the Kitt Peak solar observatory. From this data we obtained the
magnetic vector field and polarity distribution at different
heights. By a comparison of these patterns with an EIT image of Fe
XII coronal bright points, we found that the coincidence of neutral
lines and bright points at a height of about 5 Mm is highly
significant. And this feature remains still obvious up to more than
10 Mm for some bright points. Above 20 Mm, there is no clear
coincidence between bright points and neutral lines.

By comparing the projections of the upper segments of the 3-D
magnetic field lines above different heights on to the tangent x-y
plane with the image of the Fe XII coronal bright points, we come to
the conclusion that a bright point is an emission region located
around the apex of the associated magnetic loop.

From correlations between the horizontal component of the magnetic
field vector at different heights and the square root of the Fe XII
intensity enhancement in some large bright point regions, we derived
a rough height range for the possible location of each bright point.
For almost all the bright point regions analyzed, we find that with
increasing height the correlation coefficient increases to a maximum
and then decreases again. The corresponding height is defined as
correlation height, which is found to be smaller than 20 Mm for most
of the bright points. For each selected bright point, we also
obtained a height range in which the coefficient is above 95\% of
its maximum. We think the source of the bright-point emission is
probably confined within this height range. Our results indicate
that for most bright points the emission occurs below 20 Mm.

\section*{\large Acknowledgments}

This work was supported by the National Natural Science Foundation
of China under contracts 40574078, 40336053 and 40436015. It was
also supported by the Beijing Education Project XK100010404. We
thank the SOHO/EIT experiment team for providing the coronal images.
We also thank the NSO/Kitt Peak observatory for the use of their
magnetic field data.

\begin{table*}[h, t, b, p]
\caption{Correlation heights and height ranges of some bright
points} \label{table1}
\begin{tabular}{|c|p{2.5cm}|p{3cm}|p{2.5cm}|p{2.5cm}|p{4.5cm}|}
\hline Bright point & Number of data points in the bright point
region & Maximum correlation coefficient & Critical correlation
coefficient with 95\% confidence & Correlation height (Mm) &
Height range in which the coefficient is above 95\% of its maximum (Mm)\\
\hline
1 & 264 & 0.65 & 0.120 & 5 & 3.99-7.85\\
\hline
2 &  441 & 0.77  &  0.093  & 9  & 2.20-14.11\\
\hline
3 &  263 & 0.61 & 0.120 & 13 & 5.56-16.46\\
\hline
4 &  310 & 0.49  &  0.111  & 24 & 20.32-31.00\\
\hline
5 &  262 & 0.39  &  0.120  & 1  & 0.19-1.47\\
\hline
6 &  489 & 0.72  &  0.088  & 11 & 5.91-19.21\\
\hline
7 & 283 & 0.31   & 0.116  & 1  & 0-1.03\\
\hline
8 & 205 & 0.47   & 0.136  & 25 & 22.59-29.08\\
\hline
9 & 434 & 0.73   & 0.094  & 3  & 0.77-14.92\\
\hline
10 & 371 & 0.55   & 0.101  & 10 & 6.30-13.73\\
\hline
\end{tabular}
\end{table*}


\begin{thebibliography}{}

\bibitem[Ahmad \& Webb(1978)]{Ahmad78}
Ahmad, I. A., and Webb, D. F., X-ray analysis of a polar plume, Sol.
Phys., 58, 323, 1978.

\bibitem[Braj\v{s}a et al.(2004)]{Braj04}
Braj\v{s}a, R., W\"{o}hl, H., Vr\v{s}nak, B., Ru\v{z}djak, V.,
Clette, F., Hochedez, J.-F., and Ro\v{s}a, D., Height correction in
the measurement of solar differential rotation determined by coronal
bright points, A\&A, 414, 707, 2004.

\bibitem[Brown et al.(2001)]{Brown01}
Brown, D. S., Parnell, C. E., DeLuca, E. E., Golub L., and Mcmullen,
R. A., The magnetic structure of a coronal X-ray bright points, Sol.
Phys., 201, 305, 2001.

\bibitem[B\"{u}chner et al.(2004a)]{Buchner04a}
B\"{u}chner, J., Nikutowski, B., and Otto, A., Coronal heating by
transition region reconnection, J. and Walsh, R. (eds.),
\emph{Coronal Heating, Proceedings of the SOHO-15}, Vol. ESA SP-575,
2004a.

\bibitem[B\"{u}chner et al.(2004b)]{Buchner04b}
B\"{u}chner, J., Nikutowski, B., and Otto, A., Magnetic coupling of
photosphere and corona: MHD simulation for multi-wavelength
observations, Stepanov, A., Benevolenskaya, E. and Kosovichev, A.
(eds.), \emph{Multi-Wavelength Investigations of Solar Activity,
Proceedings of the IAU Symposium}, Vol. 223., 2004b.

\bibitem[Falconer et al.(1998)]{Falconer98}
Falconer, D. A., Moore, R. L., Porter, J. G., and Hathaway, D. H.,
Network coronal bright points: coronal heating concentrations found
in the solar magnetic network, ApJ, 501, 386, 1998.

\bibitem[Falconer et al.(2003)]{Falconer03}
Falconer, D. A., Moore, R. L., Porter, J. G., and Hathaway, D. H. ,
Solar coronal heating and the magnetic flux content of the network,
ApJ, 593, 549, 2003.

\bibitem[Golub et al.(1974)]{Golub74}
Golub, L., Krieger, A. S., Silk, J. K., Timothy, A. F., and Vaiana,
G. S., Solar X-Ray Bright Points, ApJ, 189, L93, 1974.

\bibitem[Madjarska et al.(2003)]{Madjarska03}
Madjarska, M. S., Doyle, J. G., Teriaca, L., and Banerjee, D., An
EUV Bright Point as seen by SUMER, CDS, MDI and EIT on-board SoHO,
A\&A, 398, 775, 2003.

\bibitem[Marsch et al.(2006)]{Marsch06}
Marsch, E., Zhou, G.-Q., He, J.-S. and Tu, C.-Y., Magnetic structure
of the solar transition region as observed in various ultraviolet
lines emitted at different temperatures, A\&A, 457, 699, 2006.

\bibitem[Masuda et al.(1994)]{Masuda94}
Masuda, S., Kosugi, T., Tsuneta, S., Hara, H., Ogawara, Y., Loop-top
hard X-ray source in a compact solar flare as evidence for magnetic
reconnection, Nature, 371, 495, 1994.

\bibitem[Metcalf et al.(1995)]{Metcalf95}
Metcalf, T. R., Jiao, L., McClymont, A. N., Canfield, R. C., Is the
solar chromospheric magnetic field force-free? ApJ, 439, 474, 1995.

\bibitem[Moore et al.(1999)]{Moore99}
Moore, R. L.,Falconer, D. A., Porter, J. G., and Suess, S. T., On
heating the Sun's corona by magnetic explosions: Feasibility in
active regions and prospects for quiet regions and coronal holes,
ApJ, 526, 505, 1999.

\bibitem[Moses et al.(1994)]{Moses94}
Moses, D., Cook, J. W., Bartoe, J.-D. F., et al., Solar fine scale
structures in the corona, transition region, and lower atmosphere,
ApJ, 430, 913, 1994.

\bibitem[Neukirch et al.(1997)]{Neukirch97}
Neukirch, N., Dreher, J., and Birk, G. T., Three dimensional
simulation studies on bright points in the solar corona, Adv. Space
Res., 19, 1861, 1997.

\bibitem[Parnell et al.(1994)]{Parnell94}
Parnell, C. E., Priest, Eric R., Titov, V. S., A model for X-ray
bright points due to unequal cancelling flux sources, Sol. Phys.,
153, 217, 1994.

\bibitem[Priest et al.(1994)]{Priest94}
Priest, E. R., Parnell, C. E., Martin, S. F., A converging flux
model of an X-ray bright point and an associated canceling magnetic
feature, ApJ, 427, 459, 1994.

\bibitem[Seehafer(1978)]{Seehafer78}
Seehafer, N., Determination of constant ¦Á force-free solar magnetic
fields from magnetograph data, Sol. Phys., 58, 215, 1978.

\bibitem[Tu et al.(2005a)]{Tu05a}
Tu, C.-Y., Zhou, C., Marsch, E., Wilhelm, K., Zhao, L., Xia, L.-D.,
and Wang, J.-X., Correlation heights of the sources of solar
ultraviolet emission lines, ApJ, 624, L133, 2005a.

\bibitem[Tu et al.(2005b)]{Tu05b}
Tu, C.-Y., Zhou, C., Marsch, E., Xia, L.-D., Zhao, L., Wang, J.-X.,
and Wilhelm, K., Solar Wind origin in Coronal Funnels, Science, 308,
519, 2005b.

\bibitem[Ugarte-Urra et al.(2004)]{Ugarte04}
Ugarte-Urra, I., Doyle, J. G., Madjarska, M. S., and
O$^{\prime}$Shea, E., Signature of oscillations in coronal bright
points, A\&A, 418, 313, 2004.

\bibitem[Vaiana et al.(1970)]{Vaiana70}
Vaiana, G. S., Krieger, A. S., Van Speybroeck, L. P., and
Zehnpfennig, T., Bull. Am. Phys. Soc., 15, 611, 1970.

\bibitem[Von Rekowski et al.(2006)]{Von06}
Von Rekowski B., Parnell, C. E., Priest, E. R., Solar coronal
heating by magnetic cancellation - I. Connection equal bipoles,
Monthly notices of the royal astronomical society, 366, 125, 2006.

\bibitem[Webb et al.(1993)]{Webb93}
Webb, D. F., Martin, S. F., Moses, D., and Harvey, J. W., The
correspondence between x-ray bright points and evolving magnetic
features in the quiet sun, Sol. Phys., 144, 15, 1993.

\bibitem[Wiegelmann \& Neukirch(2002)]{Wiegelmann02}
Wiegelmann, T. and Neukirch, T., Including stereoscopic information
in the reconstruction of coronal magnetic fields, Sol. Phys., 208,
233, 2002.

\bibitem[Wilhelm et al.(2000)]{Wilhelm00}
Wilhelm, K., Dammasch, I. E., Marsch, E., Hassler, D. M., On the
source regions of the fast solar wind in polar coronal holes   ,
A\&A, 353£¬749, 2000.

\bibitem[Xia et al.(2003)]{Xia03}
Xia, L. D., Marsch, E. and Curdt, W., On the outflow in an
equatorial coronal hole, A\&A, 399, L5, 2003.

\bibitem[Zhang et al.(2001)]{Zhang01}
Zhang, J., Kundu, M. R., and White, S. M., Spatial distribution and
temporal evolution of coronal bright points, Sol. Phys., 198, 347,
2001.

\end{thebibliography}
\end{document}